# Characterizing the Quantum Confined Stark Effect in Semiconductor Quantum Dots and Nanorods for Single-Molecule Electrophysiology

*Yung Kuo[1], Jack Li[1], Xavier Michalet[1], Alexey Chizhik[2], Noga Meir[3], Omri Bar-Elli[3], Emory Chan[4], Dan Oron[3], Joerg Enderlein[2], Shimon Weiss[1,5,6,7] \**

1 Department of Chemistry and Biochemistry, University of California Los Angeles, Los Angeles, CA 90095

2 III. Institute of Physics–Biophysics, Georg-August-Universität, 37077 Göttingen, Germany

3 Weizmann Institute of Science, Rehovot 76100, Israel

4 The Molecular Foundry, Lawrence Berkeley National Laboratory, Berkeley, California, 94720

5 California NanoSystems Institute, University of California Los Angeles, Los Angeles, CA 90095

6 Department of Physiology, University of California Los Angeles, Los Angeles, CA 90095

7 Department of Physics, Institute for Nanotechnology and Advanced Materials, Bar-Ilan University, Ramat-Gan, 52900, Israel

KEYWORDS Quantum confined Stark effect, quantum dot, membrane potential, single molecule, voltage sensor, nanorod

**ABSTRACT** We optimized the performance of quantum confined Stark effect (QCSE)-based voltage nanosensors. A high-throughput approach for single-particle QCSE characterization was developed and utilized to screen a library of such nanosensors. Type-II ZnSe/CdS seeded nanorods were found to have the best performance among the different nanosensors evaluated in this work. The degree of correlation between intensity changes and spectral changes of the exciton's emission under applied field was characterized. An upper limit for the temporal response of individual ZnSe/CdS nanorods to voltage modulation was characterized by high-throughput, high-temporal resolution intensity measurements using a novel photon-counting camera. The measured 3.5 µs response time is limited by the voltage modulation electronics and represents ~ x 30 times higher bandwidth than needed for recording an action potential in a neuron.

Nanoparticles (NPs) with extremely bright fluorescence have found application in many biological and biophysical studies[1, 2], such as single-particle imaging and tracking[3, 4], single-particle sensors[2, 5], and super-resolution microscopy[6, 7]. In particular, semiconductor quantum dots (QDs) have been used as biological probes in various applications, for example as labeling agents for long-term tracking of single molecules[3] and as intracellular sensors for temperature[5, 8], pH[9], etc. Within the past decade, QDs or nanorods (NRs) were also predicted theoretically and observed experimentally to report cellular membrane potential with exceptionally large sensitivities[10-14]. With the versatile capabilities already shown and with some advanced engineering, QDs /NRs have great potential to become next-generation voltage nanosensors that enables membrane potential imaging using single particles. In this work, we synthesized, characterized, and optimized QDs and NRs to exhibit enhanced quantum confined Stark effect (QCSE) using a dedicated high-throughput single-particle screening approach.

As neurons transmit signals via electrical impulses generated by membrane potential modulations, two approaches are traditionally taken to monitor neural activities: (i) direct electrical recording, or (ii) utilization of a sensor that transduces the electrical observables into other signals, such as fluorescence. Electrical monitoring involves the patch clamp technique or microelectrode arrays, which are both highly invasive and not suitable for simultaneous recording from the large number of neuronal cells needed in order to study how neurons communicate within a network. These techniques are also inadequate to record signals from multiple nanostructures, such as synapses or dendritic spines, within a neuron due to space constraints. Many important brain functions such as transmission of signals, plasticity, learning, and memory[15-18] are facilitated by complex electrical and chemical events in microscopic structures such as dendritic spines and synapses that volumes smaller than 1 femtoliter[17] and have extremely dynamic shapes and sizes[19-

[21]. Studying local electrical signals from such small volumes requires the development of bright and non-invasive probes that have molecular dimensions (nm), thus allowing for high spatial resolution recording. Optical sensors such as voltage-sensitive dyes (VSDs)[22-25] and genetically encoded voltage indicators (GEVIs) [26-28] are examples of the second approach. VSDs and GEVIs can report on changes in membrane potential via changes in their absorption/emission properties[29]. Despite the advances made with VSDs, GEVIs and their hybrids[29-33], they cannot monitor electrical events on the nanoscale with single molecules due to limited brightness. Additionally, most VSDs and GEVIs suffer from photobleaching, low voltage sensitivity, toxicity, and/or slow kinetics, and they can perturb membrane capacitance at high concentrations.

While organic dyes and fluorescent proteins may be further improved in the future, nanoparticles (NPs) could offer alternatives. For example, nitrogen-vacancy centers in nanodiamonds could report action potentials (APs) in a giant axon of *M. infundibulum* (worm) via optical detection of magnetic resonance (ODMR) technique[34]. QDs can operate as voltage sensors[35] via photo-induced electron transfer[12, 36, 37] or the quantum confined Stark effect (QCSE)[10, 11, 14]. The physical origin of the QCSE lies in the separation of the confined photo-excited charges in the semiconductor QD or NR, creating a dipole that opposes the external electric field. This, in turn, leads to (red or blue) shifting of the absorption and emission edges that are accompanied by quantum yield (QY) and fluorescence lifetime changes, according to the Fermi golden rule. Therefore, QDs or NRs with sizable QCSE (at room temperature) that are properly inserted into the cell membrane could report on changes in membrane potential via a spectral shift, a change in the emission intensity, and/or a change in the excited state lifetime. QCSE-based QDs/NRs voltage sensors could also offer several advantages over existing voltage sensors based on organic dyes, fluorescent proteins, or their hybrid: they (1) have high voltage sensitivity (quantified as percent change in fluorescence intensity, ΔF/F), (2) exhibit large spectral shifts (Δλ) enabling ratiometric detection, (3) exhibit changes in excited state lifetime (providing alternative detection scheme), (4) have high brightness affording single-particle detection and superresolution recording, (5) have a fast response time (~ns) based on QCSE, (6) have highly functionalizable surface, (7) have emission wavelength and quantum yield (QY) that can be engineered, (8) have negligible photobleaching, and (9) have low cytotoxicity (after surface modification). With the extremely fast response time, QDs /NRs will be capable of reporting and resolving the APs, which not only have fast dynamics (sub-ms) but also present in a wide range of frequencies and waveforms, especially in mammalian brains[38-43].

However, there are some challenges to overcome before QCSE-based QDs /NRs voltage sensors can achieve single-particle voltage imaging. First, the larger the QDs or the longer the NRs are, the larger is their polarizability and hence their voltage sensitivities[10, 13]. However, for proper membrane potential reporting, the QDs /NRs need to be functionalized and inserted into the 4 nm thick cell membrane. Therefore, a trade-off between magnitude of the QCSE (increasing with NP size and dependent on materials) and ease of membrane insertion (decreasing with increasing NP size) is required. Previous work has shown that small (~3 nm) spherical QDs can be encapsulated into liposomes[44] and short (< 10 nm) NRs, functionalized with transmembrane peptides, can be inserted into both synthetic and cellular membranes[14]. In this work, we characterize QDs/NRs with different materials, sizes, and bandgap alignments and assess the possibility of maximizing the voltage sensitivity of QDs/NRs with material and bandgap engineering without increasing their sizes. Secondly, QDs/NRs obtained by colloidal synthesis are inevitably polydispersed, especially when multiple materials or anisotropic growth (for NRs) are involved. The polydispersity among particles will introduce a large distribution in their voltage sensitivities, resulting in difficulties to evaluate the single-particle performance of these nanosensors using ensemble spectroscopic methods. These challenges need to be addressed *in vitro* before attempting to utilize QDs/NRs as single-particle voltage nanosensors in live neurons.

In this work, we addressed these challenges in several steps. First, we synthesized type-II ZnSe seeded CdS NRs (Fig. 1) with enhanced QCSE while maintaining small dimensions. A type-II heterostructure increases charge separation by presenting spatially separated band energy minima for electrons and holes across the heterojunction and hence increases voltage sensitivity by QCSE without increasing the dimensions of NRs significantly. In addition, the dot-in-rod structure of the NRs, schematically shown in Fig. 1b, is often asymmetric, creating an asymmetric charge separation and a linear dipole that screens the external field[45]. In previous theoretical work, type-II NRs were predicted to exhibit large voltage sensitivities[12, 13], and NRs < 12 nm were shown to be easily inserted in membranes after surface functionalization[14]. To address the second challenge, we developed a single-particle high-throughput screening assay, which enables iterative optimization of the synthesis. The screening was based on single-particle spectral shift ($\Delta\lambda$) and relative intensity change ($\Delta F/F$) observables and allowed for statistical evaluation of the properties of individual NRs randomly oriented in the electric field. This approach required (i) the fabrication of thin film microelectrodes that "sandwiched" NRs or QDs to allow application of homogeneous electric field vertically; (ii) building a spectrally resolved wide-field single molecule microscope; (iii) implementing a code for automated data analysis. With this high-throughput *in vitro* screening approach, we compared the voltage sensitivities and screened various QDs and NRs of different material compositions (including doping), sizes, and band alignments and acquired sufficient single-particle statistics for their assessment. Short type-II NRs exhibited larger single-particle voltage sensitivity than longer type-I NRs, validating the concept of QCSE optimization by proper engineering of NP's composition and bandgap /hetero-interface alignment. This single-particle screen allowed us to systematically and iteratively improve and optimize the NRs synthesis, and improve homogeneity, quantum yield, and voltage sensitivity. The improved NRs exhibited voltage sensitivity, characterized by a up to 69% relative intensity change ($\Delta F/F$) and a ~ 4.3 nm spectral shift ($\Delta\lambda$), fast response time (< 3.5 μs, an upper bound set by the RC time constant of the voltage modulation electronics), and high brightness, affording facile single-particle detection.

**Results:**

*Nanorod synthesis*

Reaction parameters that allow precise control of anisotropic growth are critical for NRs of small size (~10 nm) as studied here. Poor control would yield NRs with low aspect ratio (hence more symmetric and with reduced QCSE).

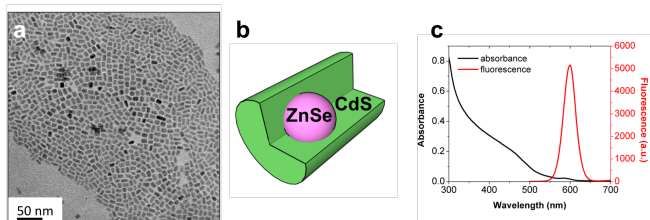

**Fig. 1:** ZnSe/CdS NRs synthesized on WANDA. (a) TEM image of NRs. (b) Schematic of the asymmetric ZnSe/CdS dot-in-rod structure. (c) Absorption and emission spectra of NR (in toluene).

Previous work reporting the synthesis of seeded ZnSe/CdS NRs focused on much longer NRs (40 nm to 100 nm)[45, 46]. Simply quenching the reaction after a short amount of time usually yielded polydisperse distributions in the length and width of the NRs. For voltage sensing, shorter NRs are required due to the difficulty to insert longer NRs (> 12 nm) into the membrane[14]. In order to determine optimal synthesis parameters to obtain shorter NRs with high aspect ratio, and to precisely control reaction parameters, we used a high-throughput robot (WANDA[47]) to systematically vary nanocrystal growth parameters and screen for NRs with aspect ratios greater than 1.5 and lengths less than 12 nm. Our ~ 12 nm NR synthesis was based on a published procedure[45] with one modification: we used an alkanethiol in the place of trioctylphosphine sulfide (TOPS) as the S precursor to control the reaction kinetics and to slow down initial growth rate. Thiol precursors of different carbon chain length (from dodecanethiol to octadecanethiol) were tested. The injection and growth temperatures respectively were optimized in multiple optimization runs, in order to achieve slower initial growth rate and hence obtain shorter NRs. The transmission electron microscopy (TEM) image and absorption and emission spectra of the batch of ZnSe/CdS NRs with minimal size distribution and high aspect ratios revealed by TEM are shown in Fig. 1. The synthesis was iteratively optimized using a series of screening assays including UV-VIS, TEM, and QCSE measurements (details shown below) to yield NRs with length < 12 nm and aspect ratio > 1.5 while maintaining a large QCSE responses. Particle size analysis of the best batch shown in Fig. 1 reports an average of 11.6 nm ± 1.7 nm for the long NR's axis, 6.8 nm ± 1.3 nm for the short axis (diameter) and an average aspect ratio of 1.9 ± 0.5 (see Fig. S1).

*Set-up and protocol for high-throughput single-particle QCSE measurements*

To characterize NRs' QCSE-induced spectral shift and intensity change upon external field modulation with sufficient statistical significance, a dedicated set-up, methodology, and analysis had to be developed. To apply electric field of magnitude similar to that generated by the depolarization phase of an AP, we followed the approach of a previous work[36] (with a few modifications) to fabricate microelectrode stacks using thin film depositions. The thin film electrodes "sandwiched" the NPs of interests and allowed application of homogeneous electric field while allowing imaging with a high numerical aperture (N.A.) objective. The stack was comprised, from bottom to top, of an indium tin oxide (ITO)-coated coverslip, $SiO_2$ for insulation,

NPs of interest, polyvinylpyrrolidone (PVP), SiO$_2$ for insulation, and gold electrode (schematically shown in Fig. S2). The NPs in such devices were stable for several weeks and up to 2 months after fabrication. NPs in older devices, such as lithographically-patterned electrodes, tended to rapidly photobleach, possibly due to oxidation by exposure to air. Compared to the horizontally patterned electrodes, the vertical stack device suffers significantly fewer catastrophic arc discharge or meltdown events[11]. This "sandwich device" can be fabricated with either a large NP density (for ensemble measurements) or with a smaller density (for single-particle measurements) by controlling the dilution and spin-coating conditions. We used low concentration to test and optimize our NRs as single-particle voltage nanosensors. A dedicated wide-field and spectrally-resolved single-molecule microscope (inspired by previous work[48], Fig. S3) was designed (see Materials and Methods and Supporting Information for details). It allowed us to acquire modulated spectra of hundreds of individual single NPs per measurement while applying voltage alternated from frame to frame. Briefly, the camera acquisition was synchronized to a modulated voltage source that was applied to the test device, creating an electric field alternating from 0 kV/cm ($V_{off}$) to 400 kV/cm ($V_{on}$) in consecutive frames. The fluorescence emitted by individual NRs were spectrally dispersed by an Amici prism inserted in the detection path and imaged by a camera. Movies of 600 frames (300 modulations periods. Frame rate at 16 Hz) were recorded for all samples.

To analyze the QCSE of single NRs, the spectrally dispersed point spread functions (PSFs) of individual NRs were selected from the mean frame of the entire movie using an algorithm for automatic PSF detection (described in Supporting Information-5), and the following analysis was performed only on the selected NRs. Wavelengths calibration for each pixel was performed according to the protocol described in Supporting Information-4, allowing the accurate extraction of the (calibrated) spectrum for each NR, anywhere in the field-of-view. Thus, the spectrum, the integrated intensity, and the peak wavelength for each NR in each frame could be automatically derived.

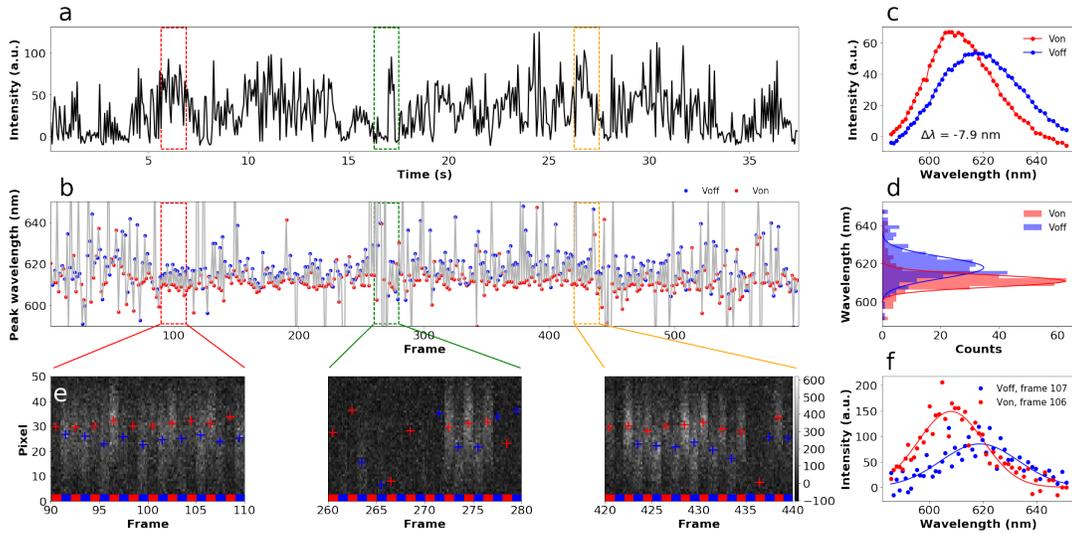

**Fig. 2:** An example for QCSE measurements of a single ZnSe/CdS NR. (a) Intensity trace and (b) spectral peak position trace as function of frames. (c) Averaged emission spectrum for $V_{on}$ (red) and $V_{off}$ (blue) frames. (d) Histograms of spectral peak positions from individual frames for $V_{on}$ (red) and $V_{off}$ (blue) frames. Fitting the two peak position distributions to Gaussian functions (red and blue lines for $V_{on}$ and $V_{off}$ histograms, respectively) showed that the centers were shifted by 7.5 nm, consistent with the result in (c). (e) Images of the prism-dispersed PSFs from 20 consecutive frames ('zoom-ins') at 3 different time periods of the background subtracted movie. The prism-dispersed PSFs from different frames are stacked side by side, with the blue and red segments marking the $V_{off}$ and $V_{on}$ frames, respectively. The peak positions are marked with red and blue +'s for $V_{on}$ and $V_{off}$ frames, respectively. The clear 'up' and 'down' displacements between consecutive dispersed PSFs are strong evidences for spectral shifts (converted to nm by calibration procedure detailed in Supporting Information-4). The intensity of each pixel in (e) is shown in greyscale with the colorbar shown on the right. (f) an example of a $V_{on}$ (red) and a $V_{off}$ (blue) spectrum from one single frame and the Gaussian fits of both spectra (blue and red lines). The noisy spectra show that finding peak positions from single frames by fitting is not ideal, and hence why, in this work, the peak positions are calculated using eq. (S3).

An example of a single ZnSe/CdS NR QCSE analysis is shown in Fig. 2. The intensity time trace and extracted spectral peak positions of a single NR are shown in Fig. 2a and 2b, respectively. The red and blue dots in Fig. 2b respectively highlight the spectral peak positions during frames with ($V_{on}$) or without ($V_{off}$) applied voltage. Clear fluctuations of intensity or peak wavelength from frame to frame due to voltage modulation can be observed. While some NRs showed clear 'on' and 'off' blinking states that can be discriminated by an intensity threshold, this NR is an example of the sub-population that does not show clear 'on' and 'off' blinking states (the histogram of total

- 10 -

counts from individual frames is shown in Fig. S5-2a). Therefore, a "burst analysis" (details described in Supporting Information-5) instead of applying a blinking threshold was implemented for further analysis to avoid the associated bias. For this NR, the averaged emission spectrum at 400 kV/cm is blue-shifted by 7.9 nm (Fig. 2c), the FWHM of the emission spectrum is narrowed by 6 nm (Fig. S5-2b), and the integrated intensity is increased by 7% (ΔF/F) with respect to the spectrum in the absence of applied field. This blue shift and increased intensity are consistent with the NR's predicted type-II band structure[11,13]. To get a preliminary estimation of NPs' voltage sensitivities, we implemented an automatic algorithm to calculate the averaged emission intensity change (ΔF/F) and spectral shifts (Δλ) (details described in Supporting Information-5) using equations below:

$$\langle \Delta F/F \rangle = \frac{\langle F_j^{V_{on}} \rangle_{F_j > th} - \langle F_j^{V_{off}} \rangle_{F_j > th}}{\langle F_j^{V_{off}} \rangle_{F_j > th}} \qquad \text{eq. (1)}$$

$$\langle \Delta \lambda \rangle = \langle \lambda_j^{V_{on}} \rangle_{F_j > th} - \langle \lambda_j^{V_{off}} \rangle_{F_j > th} \qquad \text{eq. (2)}$$

where F is the fluorescence intensity and λ is the peak wavelength. $V_{on}$ and $V_{off}$ denote the frames with ($V_{on}$) or without ($V_{off}$) applied voltage, respectively, and *th* notes the blinking threshold. The histograms of single-particle ΔF/F and Δλ, calculated using eq. (1) and (2), for all the NPs listed in Fig. 3 are shown in Fig. S5-3. In Fig. S5-3, we found that the majority (~ 80%) of type-II NRs (sample(v)) have voltage response (n=190), and the magnitudes of these averaged voltage responses, expressed in $\Delta F/F$ and $\Delta\lambda$, are much larger than that of type-I QDs (sample(i)).

**Fig. 3** Composition, band diagram and TEM images of NPs investigated in this study.

Clear spectral shifts between consecutive frames due to field modulation can also be seen in this NR. In Fig. 2e, the spectrally dispersed PSFs between consecutive frames show clear displacements, which are converted to wavelength shifts as described in SI-5. Clear blinking 'off' states are also seen in the frames missing clear dispersed PSFs (for example frame 260-271 in the second panel). The histograms of spectral peak positions (Fig. 2d) from individual frames, calculated using eq. (S3), show a large separation of 7.5 nm between the $V_{on}$ and $V_{off}$ frames, which is consistent with the peak shift found in the averaged spectra. The clear separation between the two histograms of peak positions from single $V_{on}$ and $V_{off}$ frames also show that the spectral shift can be detected from single frames without averaging.

As seen in Fig. 2, the fluorescence from single NP exhibited intermittency (blinking), and in the blinking 'on' states, there is a wide distribution in fluorescence intensity (F) and wavelength ($\lambda$) even in the absence of voltage (see Fig. 2a and 2b for examples), possibly due to local charge accumulations, ionization, and spectral diffusion. As a result, the percent intensity change ($\Delta F/F$) and the spectral shifts ($\Delta\lambda$) by voltage modulation also have a wide distribution between different modulation periods. For all NPs, $\Delta\lambda$ and $\Delta F/F$ are often not constant throughout the acquisition period but rather appears as "bursts" of large responses following periods of small or noisy responses. Moreover, averaging the $\Delta\lambda$ and $\Delta F/F$ from only blinking 'on' states requires introducing a blinking threshold, which inevitably creates bias when the 'on' and 'off' states are not clearly separated as shown in Fig. 2a and Fig. S5-2a. Instead, "burst analysis" extracts large responses, $\Delta F/F$ or $\Delta\lambda$, from time windows of 8 or more consecutive frames without applying a binary discriminating threshold. Details for the burst search algorithm are given in Supporting

Information-5. Briefly, the $|\Delta F/F|^2$ and the $|\Delta\lambda|^2$ trace (i.e. the "*score²*" trace) were first calculated between neighboring frames. A "burst" was identified when 8 (time threshold) or more consecutive frames had an averaged $|\Delta F/F|^2$ or $|\Delta\lambda|^2$ larger than the score threshold, defined as the 50% percentiles of the moving averaged "*score²*" trace with a window of 8 frames. The correlation between the Δλ and ΔF/F were small, therefore the burst search was applied separately to Δλ and ΔF/F traces. The reasons for small correlations between Δλ and ΔF/F are further studies and discussed below. Example ΔF/F and Δλ traces from the same particle with identified bursts are shown in Figs. S5-5 and S5-6, respectively. After applying the burst search algorithm to all traces from all NPs, the histograms of Δλ and ΔF/F of individual "bursts" per NP type are plotted in Fig. 4.

*Single-particle QCSE results*

With the setup and analysis for high-throughput QCSE measurements described above, we measured QCSE responses of 5 different types of NPs and build single-particle histograms to compare the voltage sensitivities, Δλ and ΔF/F, between NPs with different shapes, sizes and band alignments. Fig. 3 summarizes the properties of the NPs measured in this work, including: (i) 6 nm type-I CdSe/ZnS QDs; (ii) 12 nm quasi-type-I CdS/CdSe/CdS QDs[49]; (iii) 40 nm quasi-type-I Te-doped CdSe/CdS NRs; (iv) 40 nm quasi-type-I CdSe/CdS NRs; and (v) 12 nm type-II ZnSe/CdS NRs. Fig. 4 shows the histogram of single-particle voltage sensitivities (Δλ and ΔF/F from individual "bursts") measured from these samples. Sample (i) is the typical type-I QDs that are commercially available, in which both the excited hole and electron reside in the CdSe core. Sample (ii) is a core-shell-shell spherical QD, in which the excited hole resides in a shell shaped quantum well in the CdSe layer, and the excited electron delocalizes across the entire QD. This QD was reported to have suppressed blinking and near-unity QY[49]. Sample (iii) and (iv) are quasi-type-I NRs with and without Te doping in the CdSe core, which forms a trap for excited holes[50]. The Te-doped 40 nm CdSe/CdS NRs (sample (iii)) have on average 1 atom of Te in the CdSe core.

QCSE responses of type-I and type-II NPs are clearly distinct as shown in Fig. 4. Type-I QDs exhibit mostly red shifts ($\Delta\lambda>0$) under electric field (samples (i) - (iv)) while 47 % of bursts from type-II NRs exhibit blue shifts and 53 % exhibit red shifts (samples (v), dashed line in Fig. 4a). Type-I QDs exhibit very small negative $\Delta F/F$ (samples (i) in Fig. 4b), i.e., they decrease in intensity under electric field, while 43 % of bursts from type-II NRs exhibit positive $\Delta F/F$ and 57 % exhibit negative $\Delta F/F$ (samples (v), dashed line in Fig. 4b). We note that in the case of decreased intensity by voltage modulation, the intensity counts can only decrease to 0 in theory, i.e. $\Delta F/F$ has a lower bound of -1. The bursts that showed $\Delta F/F < -1$ (4.7 % of bursts from type-II NR, dashed line in Fig. 4b) are due to the events with extremely low counts, resulting in occasional fluctuations below 0 after background subtraction. Therefore, to avoid these events from biasing the distributions of ΔF/F's and Δλ's, histograms of both ΔF/F and Δλ bursts weighted by the total intensity counts within the bursts are also plotted in Fig. 4. Bursts with higher intensity counts have higher signal-to-noise ratio (SNR), and therefore, weighting data points by total intensity counts puts emphasis on good SNR data points. As expected, the histograms weighted in this manner show almost no population with ΔF/F's < -1. By fitting the weighted Δλ distributions with a sum of two Gaussians, two populations centered at -4.3 nm ± 2.0 nm (1 sigma from Gaussian fitting) and 3.8 nm ± 2.2 nm were found for type-II NRs (sample (v)), and only one major population centered at 1.9 nm ± 1.0 nm was found for type-I QDs (sample (i)) (gray lines in Fig. 4a). For ΔF/F distributions, fitting the weighted ΔF/F histograms with a sum of two Gaussians yielded two populations centered at 69% ± 32% and -42% ± 16% for type-II NRs (sample (v)) and only one population centered at 0% ± 7% for type-I QDs (sample (i)) (gray lines in Fig. 4b).

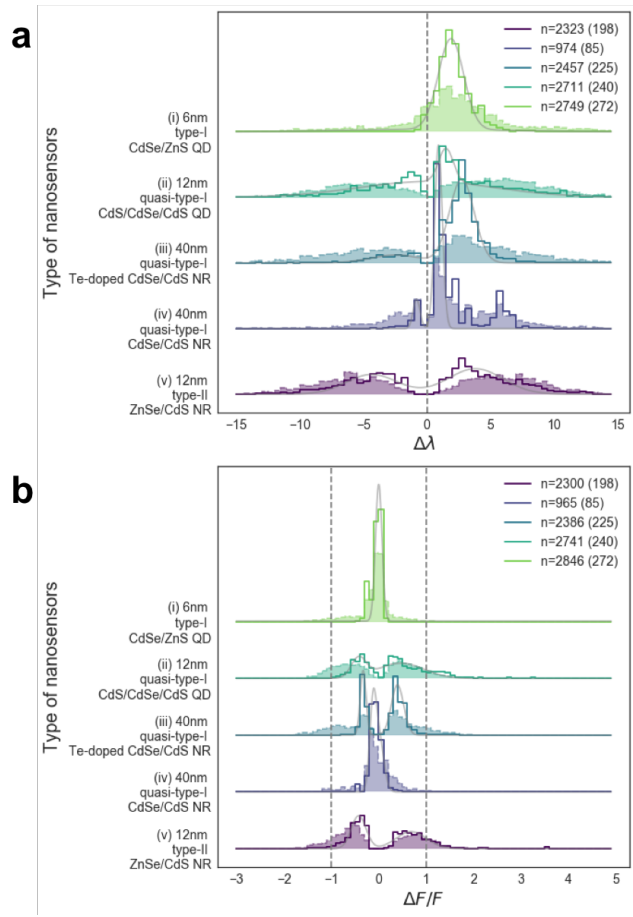

**Fig. 4:** Histograms of (a) $\Delta\lambda$ and (b) $\Delta F/F$ extracted from bursts from 5 types of NPs described in Fig. 3. The number of bursts (n) and number of NPs (number in parenthesis) in each histogram are shown in the legends. The solid lines are the histogram weighted by the total intensity counts during each burst, and the shaded areas with dashed lines are the histogram without any weight. Type-II NRs exhibit the largest voltage sensitivities, with positive and negative $\Delta F/F$ and $\Delta\lambda$ responses. The gray lines show the fits of the distributions to a sum of two Gaussians.

In summary, the type-II NRs synthesized in this work, compared to type-I QDs or 40 nm quasi-type-I NRs, exhibit both signs (positive and negative) of ΔF/F and Δλ and much larger absolute ΔF/F and Δλ for both the positive and negative populations, which are characteristics of the type-II band alignment and asymmetrically located ZnSe seeds within the CdS NRs. The type-II NRs, with largest voltage sensitivities among all NPs studies in this work and much smaller size compared to the quasi-type-I NRs, are the optimal voltage sensors among all NPs studies in this work. The average burst width, average bursts number per particle, and fractional burst duration (with respect to the entire measurement duration) are shown in Fig. S5-7. The distributions of ΔF/F and Δλ found by burst search with different time threshold was also studied and shown in Fig. S5-8. The distributions and averages of ΔF/F and Δλ are not affected by the time threshold from 8 frames to 16 frames, and the absolute values of ΔF/F and Δλ, for both positive and negative ΔF/F and Δλ, decrease as the time threshold increases to 64 frames due to averaging.

It is worth noting that in the "sandwich device", the orientations of NRs were random with respect to the direction of the applied electric field, which may broaden the distributions of voltage sensitivities. Therefore, the correlation between the QCSE spectral shifts and the orientations of the NRs was studied in a lower throughput manner using the interdigitated electrodes (Fig. 6a). Randomly oriented type-II NRs (sample (v)) were positioned between the electrodes by dropcasting, and the orientations of the emission transition dipoles of NRs were visualized using defocused imaging[51, 52]. The orientations of emission transition dipoles of the NRs were recorded without applied field, and the emission spectra were measured in the presence and absence of applied field. The results (Supporting Information-6) show 5 NRs that displayed significant spectra shifts under applied electric field. Similar results from single NRs composed of the same materials have also been shown recently[53]. In our experiment, despite the heterogeneity among the 5 type-II NRs (sample (v)) studied, a decreasing trend in the QCSE spectral shift could be seen as the angle between the transition dipole and the electric field increased from 0 degree to ~ 90 degrees. With this experiment, we found that the emission dipoles of the short type-II NRs are linear and the degree of alignment between the NR and the applied field significantly affects the QCSE spectral shift. The voltage sensitivity diminishes when the NR is orthogonal to the applied field and is maximized when the two are parallel.

To assess the amount of correlation between ΔF/F and Δλ from these type-II NRs (sample(v)), ΔF/F and Δλ of single modulation cycles and multiple cycle averages are plotted in a 2-dimensional histogram (Fig. 5a) and a scatter plot (Fig. S5-4), respectively. The correlation between ΔF/F's and Δλ's from many different particles was very small, as shown by the Pearson's correlation coefficient (-0.083 in Fig. 5a and -0.19 in Fig. S5-4). However, the correlations between ΔF/F and Δλ from single particles are much larger. The ΔF/F-Δλ Pearson's correlation coefficients were calculated for all types of NPs studied here and the results are shown in Fig. 5b. The distributions of the correlation coefficients (Fig. 5b, green) are compared to negative controls (Fig. 5b, pink), calculated as the correlation coefficients between the same ΔF/F trace and 10 randomly shuffled Δλ traces (shuffling procedure described in Supporting Information-5 and Fig. S5-10). The results show that 56% of type-II NRs (111 NRs out of 198 NRs) were negatively correlated while 20% were positively correlated with the correlation coefficients at least one standard deviation away from the mean of the controls' correlation coefficients. For type-I QDs, 43% (117 QDs out of 272 QDs) were negatively correlated while 40% were positively correlated.

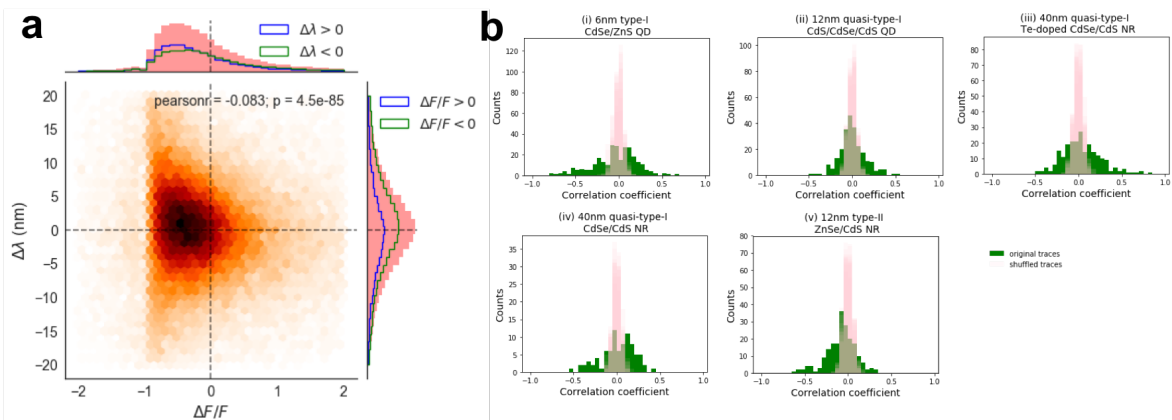

**Fig. 5:** (a) 2D histogram of ΔF/F and Δλ constructed from individual modulation cycles from 125 type-II ZnSe/CdS NRs (sample (v)) each contributing ~450 modulation cycles on average, including blinking 'off' states. The distributions of ΔF/F and Δλ are shown on the top and right panel, respectively. We note that this histogram includes all data points including blinking 'off' states and therefore is different from the distribution shown in Fig. 4, which shows only data points from "bursts". Within the top (or the right) panel, the histograms of sub-populations, Δλ (or ΔF/F) > 0 and Δλ (or ΔF/F) < 0, are plotted as the blue and the green lines, respectively. (b) Histograms of Pearson's correlation coefficients extracted from individual NPs for each type of NP (green). The distributions are compared to the distributions of Pearson's correlation coefficients calculated between the same ΔF/F traces and 10 randomly shuffled Δλ traces (pink, serving as controls).

The results show that for the majority of NPs, the ΔF/F and Δλ from the same NPs were correlated within the acquisition time, meaning that an intensity change is usually accompanying a wavelength shift with proportional magnitude as predicted by the theory of QCSE. However, for some NPs, the correlation coefficients are positive instead of negative, opposing the prediction by the theory of QCSE. Meanwhile, the correlation between ΔF/F and Δλ from different NPs show that the sign and the magnitude of ΔF/F and Δλ could not be predicted purely by Stark effect, possibly due to different rates of charge trapping and de-trapping in the defect states, different degrees of surface passivation and different local environments for different particles.

As reported previously, QDs and NRs may experience changes in emission wavelength and blinking rate or photobleach after a long period of excitation in ambient air due to photochemical reactions (oxidation) of the surface of the nanocrystals[54-57]. Therefore, the stability and voltage sensitivity of a single type-II NR (sample (v)) after a long period of excitation were also studied. The emission of a single, type-II NR was recorded using a dual-view spectral splitting setup while the camera recording was synchronized to the alternating applied electric field. The experimental detail and the results are shown in Supporting Information-6. The NR was continuously excited under ambient air, and the fluorescence was recorded for more than 8 h. The voltage sensitivities, including both ΔF/F and Δλ, despite some fluctuations due to blinking, remain nearly constant in the 'blinking on' states for more than 4 h. Over time, the emission blue shifted, and the NR entered more dark states, while the Δλ decreased after 5 hrs. Although exposed to ambient air, this single NR exhibit excellent long life (> 8 h) and voltage-sensing stability (~4 h). With improved coatings, even longer lifetimes are expected in the future.

*Temporal response of single type-II ZnSe/CdS NRs*

To capture an AP, a membrane voltage sensor needs to have a sub-ms temporal response and a photon emission rate which allows recording it with sufficient signal-to-noise ratio (SNR). In principle, the response time of NRs is on the order of their excited state lifetime, which is of the order of a few tens of nanoseconds. Bar-Elli *et al.* have shown that bright NRs (with photon emission rate of ~$10^5$ Hz) could report voltage at 1kHz via their spectral shifts, $\Delta\lambda$[53]. Here we set up to directly resolve the fluorescence intensity of individual NRs in response to voltage modulations with a custom wide-field photon counting camera (PCC) designed for time-resolved single-molecule imaging[58-62].

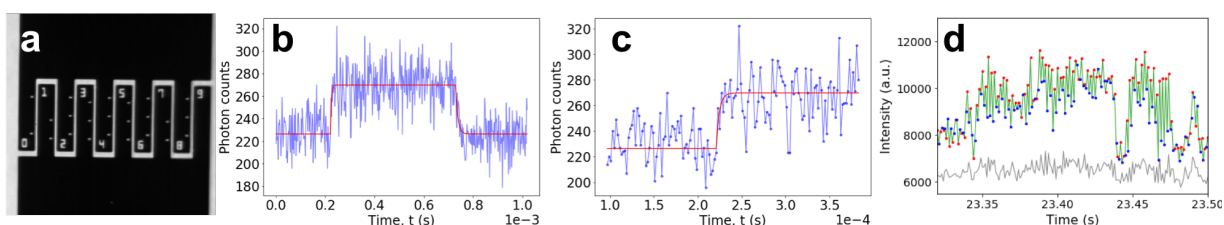

**Fig. 6:** Temporal response of a single type-II NR (sample (v)). (a) An image of the lithographically patterned microelectrode (with 2 μm gap) used in this experiment. (b) Accumulated photon counts from one single NR (from an acquisition of 269 seconds, blue) as a function of photon arrival time with respect to the voltage trigger. The time bin is 2 μs, and the modulation frequency was 1 kHz. The red line is a fit to the photon count trace with equation (1) to extract time constants. (c) Zoom-in on the rising edge in (b), from which the fit (red line) yields a response time of 3.5 μs. (d) Single type-II NR response recorded using a spectral splitting dual-view setup (details shown in Supporting Information) at 1 kHz frame rate with alternating applied voltage at 500 Hz. The green trace is the integrated emission intensity from a single NR from the two dual-view channels. The red and blue dots mark the intensity extracted from the $V_{on}$ and $V_{off}$ frames, respectively. The gray line is the sum of the backgrounds extracted locally from pixels near the NR in the two channels.

For this experiment, we deposited NRs in between lithographically patterned horizontal microelectrodes[11] with a 2 μm gap (Fig. 6a) and imaged them with the PCC while applying a modulated voltage on the electrodes. In contrast to conventional cameras which accumulate photoelectrons generated during a preset integration time, readout the whole frame at fixed interval, and suffer from readout noise, the PCC records each photon's arrival times with 156 ps temporal resolution, and its location with 50 μm spatial resolution, allowing to define arbitrary "frame" durations post acquisition and is readout-noise-free. The maximum achievable frame rate is only limited on one hand by the signal and the desired SNR, but also by detector hardware constraints. In the device used for these experiments, local (single NR) count rate was limited to

~40 kHz, while the global count rate (over the whole detector) was limited to ~2 MHz. For periodic processes such as studied here, the availability of precise photon timing allows for accumulation of data from "frames" of duration much shorter than the above limits, the signal being now limited only by the total acquisition time. This allowed us to define frame rates comparable to the fastest complementary metal oxide semiconductor (CMOS) cameras but with single-photon sensitivity due to the absence of readout noise.

Type-II NRs in toluene were drop-casted on the microelectrodes and covered with polyvinylpyrrolidone (PVP) by spin-coating after toluene evaporation. During the acquisition of photons by the PCC, 1 kHz square voltage modulation (at 50% duty cycle) was applied to create alternating electric fields alternating between 0 and 400 kV/cm. The triggers for the voltage modulations were simultaneously recorded by the PCC electronics and time stamped with an internal 50 MHz clock synchronized with that used to time-stamp detected photons. 269,904 modulation periods were recorded for the single-particle intensity time trace shown in Fig. 6b (total duration: 269 s). Fig. 6a shows the electrodes used for this experiment imaged under a microscope with white light illumination. Fig. 6b shows the accumulated photon counts after reassignment to a single modulation period. The results shown here represents an assessment of the time resolution when a measurement is not limited by the photon emission rates, for example, when an ensemble of oriented NRs responding collectively to the electric field. The ΔF/F of this specific NR was 19%.

To extract the temporal response of the NRs, we fitted the signal with the following model:

$$F(t < \mu_1) = F_0$$

$$F(\mu_1 < t < \mu_2) = \Delta F \left(1 - e^{-\frac{t-\mu_1}{\tau_1}}\right) + F_0$$

$$F(t > \mu_2) = \Delta F \left(e^{-\frac{t-\mu_2}{\tau_2}}\right) + F_0 \qquad \text{eq. (3)}$$

where $F$ is the photon counts, which is a function of time, $t$. $\mu_1$ and $\mu_2$ are the fitted start time of the NR's intensity rise and decay, respectively. $\tau_1$, and $\tau_2$ are the fitted rise and decay time constant, respectively. $F_0$ represents the total photon counts in the absence of voltage, and $\Delta F$ is the change in photon counts under applied voltage with respect to $F_0$. Therefore, $\Delta F/F$ equals $\Delta F/F_0$. Fitting the accumulated trace to eq. (3) yielded time constants $\tau_1 = 3.5$ μs and $\tau_2 = 8.8$ μs (Figs. 6b, c) for the rising and falling edges, respectively. The time constant of 3.5 μs is exactly the rise time of the measuring circuit (including the voltage amplifier and RC time constant of the microelectrodes, Figs. S8-1 and S8-2). The measured response time is therefore an upper bound imposed by the instrument, and is ~ 30 times faster than what is needed to resolve an AP. We note that, however, in order to achieve kHz recording for APs using a single NR, the emission rate of this specific NR must exceed 30 kHz to overcome the shot noise to allow for observing a 19% change in intensity measured previously. In the current setup, the global count rate was limited by the PCC, and therefore a single NR emission rate of ~ 4 kHz was recorded, which was well below the saturation emission rate of the NRs. To demonstrate that these NRs could indeed report voltage at 1 kHz frame rate with sufficient photon emission, the fluorescence from a single NR was recorded using a dual-view spectral splitting setup (details described in Supporting Information) and an EMCCD with a small field-of-view (50 x 50 pixels). The results are shown in Fig. 6d and Supporting Information-7. The NR exhibited alternating emission intensity and emission peak positions, in single 1 ms frames without averaging, as the applied electric field alternated between 0 kV/cm and 300 kV/cm from frame to frame (Fig. 6d). The result shows that the QCSE from ***a single NR*** could be observed using the spectral splitting setup at 1 kHz frame rate, which is consistent with the recently published work[53]. With proper functionalization for membrane insertion, a single NRs could therefore be potentially used for monitoring neural activities at ~ millisecond temporal resolution.

**Discussion:**

The optimal voltage sensors for single particle electrophysiology need to have large voltage sensitivity while maintaining small sizes for membrane insertion. Using a high-throughput QCSE screening assay, we successfully demonstrated that the 12 nm long type-II NRs (sample(v)) exhibit much larger voltage sensitivities (ΔF/F and Δλ) compared to the other NPs studied in this work, including spherical QDs and 40 nm long quasi-type-I NRs. Type-II band alignment increases the separation of excited holes and electrons and hence QCSE, and therefore, by changing the material compositions of seeded NRs, the voltage sensitivities (ΔF/F and Δλ) can be increased without increasing the sizes of the NRs. Although the length of our type-II NR is longer than the thickness of a lipid bilayer membrane of ~ 4 nm, a method for functionalizing similar semiconductor NRs for vertical insertion into cellular membranes has been demonstrated experimentally[14]. The voltage response of a type-II NR of similar length has also been calculated by Park *et al*[13], using self-consistent Schrödinger-Poisson calculations for NRs inserted vertically across the membrane. The result showed that a 12-nm-long type-II NR exhibits largest voltage sensitivity (ΔF/F > 190 % / 100 mV) compared to shorter type-II NRs and type-I NRs with the same length. The experimental results presented here validate our previous predictions of the optimal materials and dimensions for NR voltage sensors.

One of the important characteristic required for membrane potential sensors is that they can report the physiological membrane potential, ranging from ~-70 mV to 40 mV. Hence, a good membrane potential sensor must have asymmetric responses around zero field. In type-I spherical QDs, the excited dipole is small and isotropic due to quantum confinement in the spherical core (well), while in type-II NRs, excited carriers separate to different ends of the NRs, creating a linear dipole that partially screens the external electric field. While type-I QDs exhibit isotropic responses regardless of the field direction, the linear dipole in a NR results in asymmetric (positive and negative) ΔF/F and Δλ responses depending on the direction of the electric field with respect to that of the dipole, as well as larger absolute values of ΔF/F and Δλ under both signs[13]. The results by burst search (Fig. 4) and by comparing the Δλ distributions with that from randomized wavelength (λ) traces (Fig. S5-10) both show that type-II NRs exhibit Δλ of both signs while type-I QDs and quasi-type-I NRs exhibit primarily positive Δλ's. The presence of both blue shifting and red shifting NR populations was consistent with theoretical predictions[13] and attests to the type-II nature of these NRs and their asymmetric structure. Previous work has shown that seeded NR growth is often asymmetric and the seed is offset from the center of the NR by ~1/3 length of the NR[46] (as illustrated in Fig. 1b), resulting in spatially asymmetric distribution of carrier wavefunctions. These NRs, with both positive and negative ΔF/F and Δλ due to random orientations in the electric field, are capable of reporting not only the field strength but also the field direction, as required for electrophysiology applications.

For type-II NRs, despite some heterogeneity, the average ΔF/F is very large (69% for positive ΔF/F and -42% for negative ΔF/F, Fig. 4), and the average Δλ is +3.8 nm and -4.3 nm for an electric field of 400 kV/cm, which is approximately 1.45 times larger than the electric field swing that will be created during firing of an AP in a neuron. The ΔF/F from the type-II NRs, even when divided by a factor of 1.45, is much larger than most commercially available VSDs (di-8-ANEPPS: 7.5% per 100 mV[63]; di-4-ANEPPS: 4.13% at 560 nm and 2.12% at 620 nm per 150 mV[64]; RH237: 2-3 %[65]; JPW-6003: 11.9%[66]) or GEVIs and has much higher signal-to-background ratio at the single particle level.

One potential challenge for using these NRs as voltage sensors is that our QCSE screening results showed broad distributions of voltage sensitivities (ΔF/F and Δλ) for single type-II ZnSe/CdS NRs (sample(v)). Some NRs exhibited large spectral shifts (red shifts or blue shifts), while some exhibited small or no spectral shifts. Some NRs exhibited large changes (positive or negative) in emission intensity (ΔF/F), while some exhibited small or no changes in ΔF/F. The main reason for this is the wide distributions of random orientations of NRs with respect to the electric field direction. The experimental results showed that the misalignment between the NRs and the electric field can diminish the voltage sensitivity, and that the random orientations of NRs are responsible for the distributions of voltage sensitivities shown in Fig. 4. In the "sandwich device", the surface roughness of the substrate and the residuals, such as excess ligands, from the solution could cause the NRs to be not completely orthogonal to the vertical applied field and hence a distribution of spectral shifts offset from 0. Other possible causes of the wide distribution in voltage sensitivities are the heterogeneity (in sizes, shape, and QYs) among particles. However, with proper functionalization, we expect better control over the alignment between the NRs and the electric field in applications in lipid membranes, overcoming the issue of broad distributions of voltage sensitivities caused by random orientations of NRs as shown in this work.

Another consideration before using these NRs as voltage sensors is that, according to the theoretical studies by Park and Weiss[13], $\Delta$F/F is dependent on the QY of the NP as well as on the excitation power (which in turn affects the QY via re-partitioning between the exciton state and the positive trion state). For a 12 nm type-II ZnSe/CdS NR with 10% QY and 100% partitioning in the exciton state, ΔF/F is estimated to be > 60%, while the same NR with 90% QY and 20% partitioning in the exciton state exhibits a ΔF/F of only ~5%[13]. While a larger QY and larger excitation power could allow a higher emission rate, they also decrease ΔF/F and the partitioning in the exciton state. A careful balancing of the trade-off between ΔF/F, QY and excitation power is therefore needed in order to achieve AP recordings.

An interesting finding unexpected by Park and Weiss is that the ΔF/F responses in the colloidal NRs are more complicated than the QCSE theory could predict. The ΔF/F caused by the intrinsic QCSE effect, without considering defect or charge trapping states, is negatively correlated with the Δλ. However, all the NPs studied in this work showed a fraction of the population with positively correlated Δλ and ΔF/F. The reasons for positively correlated Δλ and ΔF/F could be that ΔF/F was induced by a combined result of the applied electric field and local charges at surface and interface defects[67] since local charge states can create a local electric field[10] and be modulated by the applied electric field[68]. Extrinsic charging/ionization at surface- and interface-defects[68] could further modulate blinking rates (and hence QY) and contribute to a positive correlation and/or no correlation. These effects will require further studies that correlate spectral, intensity and lifetime measurements under applied electric field and at different excitation powers (currently a topic of a follow-up project). Such measurements will allow us to decouple these contributions and further improve type-II NRs as voltage sensors (for example, by growing an additional layer of a high bandgap coat). Indeed, the high-throughput screening method developed in this work is most suitable for these studies and will be used to minimize extrinsic effects and optimize the intrinsic QCSE signal.

Furthermore, the large Δλ shift from the type-II NRs (relative to their emission spectral width) is amenable to ratiometric detection using a dual-channel spectral splitting setup[11, 53] as shown by the single NR recording results. Δλ-based, ratiometric measurements are also immune to the complications associate with the intensity-based measurements described above. Together with further improvements in the sensors' design and performance and in their surface functionalization for membrane insertion[14], we envision their utilization for parallel, multi-site, super-resolved electrophysiological recordings.

**METHODS**

*Materials and chemicals:* All chemicals are used as purchased without further purification. Trioctylphosphine oxide (TOPO, 99%), Octadecylphosphonic acid (ODPA) and hexylphosphonic acid (HPA) were purchased from PCI Synthesis. Tri-n-octylphosphine (TOP, 97%) was obtained from Strem Chemicals. Cadmium oxide (CdO), octadecylamine (ODA), hexadecylamine (HDA), octadecanethiol (ODT), 1.0 M diethylzinc ($Zn(Et)_2$) solution in hexanes, were purchased from Sigma-Aldrich. Selenium powder (99.999%, 200 mesh) was purchased from Alfa Aesar.

*Synthesis of ZnSe/CdS NRs:* The detailed procedure for synthesis of ZnSe QDs is described in Dorfs et al[45]. Briefly, a mixture of Se (63 mg), TOP (2 g) and diethyl zinc solution (0.8 ml, 1M) was injected into degassed HDA (7 g) at 300 ℃ in argon atmosphere. The reaction was kept at 265 ℃ until a sharp absorption peak around 360 nm was observed (~30 mins after injection). After the reaction was cooled to room temperature, ZnSe QDs were purified 3 times by butanol/methanol precipitation and redissolved in toluene. The concentration of ZnSe in toluene was documented by the optical density (OD) at the absorption peak through a 1 cm cuvette. To synthesize CdS nanorods on ZnSe seeds using WANDA, CdO (270 mg), ODPA (1305 mg), HPA (360 mg), and TOPO (13.5 g) were first degassed at 100 ℃ under vacuum for 2 hr, and the solution was heated to 230 ℃ until the CdO powder was fully dissolved, rendering a colorless solution. The solution was cooled to, room temperature to add 180 mg ODA, and the solution was degassed under vacuum at 100 ℃ for additional 2 hr. To prepare the S precursor solution with ZnSe, 1440 mg of ODT were mixed with 36 units [OD (under 1cm path length) × ml] of ZnSe solution in toluene and heated under vacuum to remove the toluene and moisture. After degassing, both Cd precursor solution and S precursor with ZnSe were transferred under vacuum into a glove box and dispensed gravimetrically into the 40 ml glass vials used as reaction vessels for the robot. The filled vials were loaded into the eight-reactor array of WANDA, an automated nanocrystal synthesis robot at the Molecular Foundry[47]. WANDA was used to run up to eight reactions in serial with individually controlled heating/cooling profiles, stirring rate, injections and aliquot schedules. Below is the description of an exemplary run. 1.133 ml of S/ZnSe solution (heated to 50 ℃ to prevent solidification) was injected into 15615 mg of Cd solution at 330 ℃ at a dispense rate of 1.5 ml/sec. The temperature after injection was set at 320 ℃ for CdS NR growth. The heating was stopped 15 mins after injection. To thermally quench the reaction, each reaction was then rapidly cooled to 50 ℃ using a stream of nitrogen, after which 5 ml of acetone was injected.

*Post-synthesis treatment of ZnSe/CdS NR*: The obtained product was purified 3 times by methanol precipitation and toluene wash/centrifugation to remove free ligands and unreacted precursors, and a half monolayer of Zn was grown on the NRs' surface to introduce metal rich surface and coordinating ligands for further functionalization while maintaining QY (30-40%) of these NRs during and after functionalization. Briefly, the purified ZnSe/CdS NRs were mixed with TOPO, oleic acid and oleylamine. Following degassing, the reaction solution was heated to 250-280℃ under argon, a TOP solution of zinc acetate or zinc undecylenate was infused to the reaction flask with a needle mounted syringe. The reaction was stopped by removing the heating mantle 20 min later.

*Wide field spectrally resolved microscopy for QCSE measurements:* A wide field microscope based on Zeiss Axiovert S100TV, with home-built illumination and detection optics, was used. A 460 nm laser (Sapphire 460-10, Coherent) was focused onto the back focal plane of a 100× objective (Zeiss Plan-Neofluar, N.A. 1.3, oil immersion) to create wide-field illumination. The laser power was 0.7 mW before entering the objective. A 488 nm dichroic mirror (Di03-R488, Semrock) and a 530 nm long pass filter (E530LP, Chroma Technologies) were used. In the detection path, a removable Amici prism could be inserted before the electron-multiplying charged couple device (EMCCD) (Ixon DU-897, Andor) for spectrally dispersing PSFs. For each QCSE measurements, a wide field image without the Amici prism was first acquired to locate each NP. After inserting the prism, a movie was acquired while synchronously alternating the voltage applied to the sandwich device as described below. Alternating voltage between 0 ($V_{off}$) and ~60 V ($V_{on}$) (variable depending on the final thickness of $SiO_2$, PVP, and $SiO_2$ layer in each device) was generated by a function generator (FG2A, Beckman Industrial) creating square wave at 8 Hz with 50% duty cycle and amplified with an additional offset of ~30 V (variable, to offset the voltage in half periods to 0 V) by a high bandwidth voltage amplifier (STM100, RHK Technology). The exact voltage applied for $V_{on}$ was calculated to impose an electric field of 400 kV/cm in the PVP layer, assuming that the two $SiO_2$ layers and the one PVP layer were acting as three capacitors in series with dielectric constants of 3.9 and 2.33 for $SiO_2$ and PVP, respectively. Synchronization of the function generator and the EMCCD camera was achieved by a programmable FPGA board (410-087, Digilent Inc.), which identified downward and upward zero voltage crossings from the voltage generator and output TTL triggers at each crossing. As a result, two frames were recorded for each modulation period (one frame with voltage on, one frame with voltage off). Therefore, the resulted exposure time for each frame was 62.5 ms, and the frame rate was 16 Hz. All movies consisted of 600 frames. The algorithms for extracting QCSE parameters from these movies is described briefly below and in detail in Supporting Information-5.

*Wide-field photon counting microscopy using a photon-counting camera:* An inverted wide field microscope based on Olympus IX71 with home-built excitation and detection optics was constructed. The excitation source was 532 nm continuous wave laser (MGL-III-532-150mW, Opto Engine). The laser was reflected by a 488/532 long pass dichroic mirror (Omega Optical, transmission spectrum shown in Supporting Information-9) and focused onto the back focal plane of a 60× objective with a numerical aperture (N.A.) of 1.45 (Plan Apo TIRFM, Olympus) to create wide field illumination. The excitation power was 1.9 mW before entering the objective. A 596/60 band pass filter (596DF60, Omega Optical) was used as the emission filter. Details for the hardware and software for the photon-counting camera (PCC) are similar to those presented in Colyer *et al*[62] with the following differences: the detector was comprised of a GaAs photocathode cooled down to 12 °C, with quantum efficiency of ~30% in the detection wavelength range and a cross-strip anode for position sensing[69]. Dilute NR solution in toluene were drop-casted onto lithographically patterned electrodes, allowed to dry in air, and covered with a layer of PVP by spin-coating. After locating NRs in between the electrodes under microscope, square wave voltage was applied via the same hardware used for QCSE measurements, while the emitted photons from NRs were detected and time-stamped by the PCC. Since the detector does not accumulate photons into "frames" but instead collects position and time information for each photon, synchronization of acquired photons was performed with the help of TTL triggers emitted by the voltage function generator, recorded as a separate time-stamped signal by the PCC electronics. Because of hardware limitation, only 1 every 8 trigger signal was recorded, which provided plenty of information for post-acquisition synchronization, each time-stamp being recorded with 20 ns resolution.

*Fabrication of sandwich devices:* An 18 mm x 18 mm indium tin oxide (ITO) coated coverslip (#1.5, 30-60 ohms per square resistivity, SPI Supplies) was used as the starting substrate. 500 nm of a SiO$_2$ layer was deposited using e-beam evaporation (Mark 40, CHA) at a rate of 1.5 Å/s. The resistance of the substrate after SiO$_2$ deposition was tested and confirmed to be infinite using a multimeter. QDs or NRs in toluene were spin-coated on top of the SiO$_2$ layer, followed by spin-coating of 5% w/w poly-vinylpyrrolidone (PVP, 40k Sigma-Aldrich) in 1:1 methanol and H$_2$O solution to create a layer of PVP of 400-500 nm (measured by a stylus profilometer, Veeco Dektak 8, Bruker). Next, a second 500 nm SiO$_2$ layer was deposited on top of the PVP layer for insulation, followed by the second electrode deposition of 5 nm Cr layer (0.1 Å/s) and a 100 nm Au layer (1 Å/s) using e-beam evaporation (Mark 40, CHA). The deposition of the second electrode was through a shadow mask that created six electrodes of 3 mm diameter on each coverslip. The thicknesses of each layer were measured with the stylus profilometer (Veeco Dektak 8, Bruker) after each deposition or spin-coating to assist calculation of the voltage required for QCSE measurements.

Supporting Information:

NR size analysis, Schematics of the "sandwich device", Schematics of the wide-field spectrally-resolved microscope, Wavelength calibration for the spectrally-resolved microscope, Data analysis for extracting QCSE results, Dependences of QCSE on the orientations of NRs with respect to the electric field and the exposure time, Single particle recording at 1 kHz, Characterization of the instrument's RC time constant, Dichroic mirror's transmission spectrum

AUTHOR INFORMATION

**Corresponding Author**

*Corresponding author. E-mail: sweiss@chem.ucla.edu**Present Addresses**

- 30 -

†If an author's address is different than the one given in the affiliation line, this information may be included here.

**Author Contributions**

Y.K. conducted all experiments, analyzed all the data, and wrote the manuscript. J.J.L. and Y.K. synthesized ZnSe/CdS NRs using WANDA. ZnSe QDs were synthesized by J.J.L.. X.M. built the optical setup with the PCC with excitation source added by Y.K.. X.M. and Y.K. performed related data analysis. A.C. perform defocused imaging experiments. N.M. synthesized the 40nm NRs. E.C. contributed to NR syntheses and provided training for WANDA operation. D.O. provided 2 types of NRs. X.M., A.C., O.B., D.O., J.E. contributed to discussions. S.W. and Y.K. designed the experiments. S.W. X.M., and D.O. helped in writing and revising the manuscript. All authors have given approval to the final version of the manuscript.

ACKNOWLEDGMENT

We would like to thank Antonio Ingargiola and Kyoungwon Park for discussion on data analysis, Max Ho and Wilson Lin for discussion on thin film fabrication, Andrew Wang and Ocean Nanotech LLC for providing the CdSe/ZnS QDs at no cost, and Prof. Wan Ki Bae for providing the CdS/CdSe/CdS QDs. This research was supported by DARPA Fund #D14PC00141, by the European Research Council (ERC) advanced grant NVS #669941, by the Human Frontier Science Program (HFSP) research grant #RGP0061/2015 and by the BER program of the Department of Energy Office of Science, grant # DE-FC03-02ER63421. Work at the Molecular Foundry was supported by the Office of Science, Office of Basic Energy Sciences, of the U.S. Department of Energy under Contract No. DE-AC02-05CH11231. This work was also supported by the National Science Foundation under Grant No. DMR-1548924

Astronomy. In *Advanced Maui Optical and Space Surveillance Technologies Conference*, Maui, HI, 2012.